\documentclass[twocolumn,showpacs,amsmath,amssymb,superscriptaddress,nofootinbib]{revtex4-1}

\usepackage{hyperref}
\usepackage{scrextend}
\usepackage{bibentry}

\usepackage{natbib}
\usepackage{color}

\usepackage{graphicx}% Include figure files
\usepackage{dcolumn}% Align table columns on decimal point
\usepackage{bm}% bold math
\usepackage{enumerate}
\usepackage{slashed}
\usepackage{simplewick}
\usepackage{comment}

\usepackage{soul}

\usepackage[english]{babel}
\usepackage{appendix}
\usepackage[utf8]{inputenc}

\usepackage{marginnote}
\usepackage[normalem]{ulem}

\newcommand{\ii}{{\rm{i}}}

\newcommand{\nn}{\nonumber}

\newcommand{\eq}[1]{(\ref{#1})}
\renewcommand{\>}{\rangle}
\newcommand{\la}{\label}
\newcommand{\ba}{\begin{align}}
\newcommand{\ee}{\end{equation}}
\newcommand{\be}{\begin{equation}}

\def\12{\frac{1}{2}}
\newcommand{\p}{\partial}
\newcommand{\en}{\end{align}}

\newcommand{\<}{\langle}

\newcommand{\mn}[1]{\marginpar{\tiny #1}}
\usepackage{color}

\begin{document}

\title{Collective Field Theory for  Quantum Hall States  }
\author{ M. Laskin}
 \affiliation{ Department of Physics, University of Chicago, 929 57th St, Chicago, IL 60637, USA}
\author{ T. Can}
\affiliation{ Simons Center for Geometry and Physics, Stony Brook University, Stony Brook, NY 11794, USA}
\author{ P. Wiegmann}
 \affiliation{ Department of Physics, University of Chicago, 929 57th St, Chicago, IL 60637, USA}%Lines 

\date{\today}

\begin{abstract}
 We develop a collective field theory for fractional quantum Hall (FQH) states. We show that in the leading approximation for a large number of particles, the  properties of Laughlin states are captured by a Gaussian free field theory with a background charge.   Gradient corrections to the Gaussian field theory arise from  the covariant ultraviolet regularization of the theory, which produces the gravitational anomaly. These corrections  are described by  a theory  closely related to the  Liouville theory of quantum gravity.   The field theory simplifies the computation of correlation functions  in FQH states and  makes manifest the effect of quantum
 anomalies.

\end{abstract}

\date{\today}

\maketitle

%73.43.-f       Quantum Hall effects
%73.43.Cd       Theory and modeling
%73.43.Lp       Collective excitations
%02.40.-k       Geometry, differential geometry, and topology

%\tableofcontents
\newpage
\paragraph[1]{ Introduction}
Since the work of Laughlin \cite{Laughlin1983}, a common approach to  analyzing the physics of the fractional  quantum Hall  effect (FQHE)  starts with  a trial ground state wave function for \(N\) electrons. Despite its success, this approach is an impractical framework for studying  the collective behavior of a large number of electrons  ($N\sim
10^6$,   in samples exhibiting the QHE). As a result, some subtle properties  of QHE states, such as   the  gravitational anomaly \cite{clw, CLWBig,Klevtsov2013, Abanov2014,CLWBig, GromovAbanov, Klevtsov2014,  framinganomaly,Read2015,KW2015}, were computed only recently.  

 The effects of {\it\ quantum anomalies} are essential in the physics of the QHE. 
Although anomalies originate at short distances on the order of the magnetic length,  they control the large-scale properties of the state, such as transport. It was recently shown in \cite{KW2015} that, like the Hall conductance, transport coefficients determined  by the gravitational anomaly are expected
to be quantized on QH plateaus. For this reason it is important to formulate
the theory of the QH  effect in a fashion which makes the quantum anomalies manifest.
 The field theory approach seems the most appropriate for this purpose.

In this paper, we develop a field theory for Laughlin states. This approach naturally captures universal features of the QHE, and emphasizes the 
geometric aspects of QH-states. We demonstrate
how the field theory encompasses recent developments in the field \cite{clw,Klevtsov2013, Abanov2014, CLWBig,GromovAbanov,Klevtsov2014, framinganomaly,Read2015,KW2015}
and obtain some properties of quasi-hole excitations.  A preliminary treatment of this approach appears in \cite{CLWBig}.

 The universal properties of the QHE are encoded in the dependence
of the  ground state wave function
on electromagnetic and gravitational backgrounds (see e.g., \cite{clw}). For that reason we study QH states on a Riemann surface and for simplicity focus on genus zero surfaces.

 We restrict our analysis to the Laughlin states.  Our   approach is closely connected to the  hydrodynamic theory of QH states of Ref \cite{Wiegmann2013,*PW13} and  the collective
field theory approach of Gervais, Sakita and Jevicki developed in  \cite{Gervais1976,*SakitaJevicki,*JevickiSakita2}  and extended   in \cite{Awata1,*Awata2,AbanovWiegmannCalogero,
*AbanovWiegmannBettelheimCalogero}.  The  action of the field theory for Laughlin states is written in Sec.(3). The leading part, Eq.\eq{gammaG}, is equivalent to the classical energy of a 2D neutralized Coulomb plasma when the discreteness of particles is not taken into account. This is used in the familiar plasma analogy of Ref.\cite{Laughlin1983} to deduce the equilibrium density, as well as properties of the quasi-hole state such as charge and statistics. The other terms in the action are more subtle but equally significant, and give rise to important effects including the gravitational anomaly. 
 \smallskip

\paragraph{ Collective Field Theory}
We start with some general remarks about the collective field theoretical approach.

%   We write the $N$-particle wave function $\Psi(\xi_{1}, ..., \xi_{N})$ %as a function of particle coordinates $\xi = (z, \bar{z})$ on a Riemann %surface with metric in conformal gauge $\sqrt{g} dz d\bar{z}$. The fact %that the configuration space is Riemannian manifold is not crucial at this %stage, but sets the scene for our later discussion. 
 To compute the expectation value of an observable $\mathcal O(z_{1},...,z_{N})$ within the ground state \(\Psi(z_1,\dots,z_N)\), one has to evaluate a multiple integral over the individual particle coordinates\begin{align}\la{2}
\< \mathcal{O}\>=\int \Psi^\ast\mathcal{O }\Psi \; dV_1\dots dV_N\, ,  \quad    dV_i = \sqrt {g(z_i)} d^2z_i,
\end{align}
 and then  proceed  with the large \(N \)  limit.
%  Here, $dV = (i/2)\sqrt{g} dz \wedge d\bar{z}$ is the volume form on the %surface.
  The field theory approach assumes instead that the appropriate  variables are collective modes. In the QH  systems the ground state at a fixed background
gauge  potential  is
a holomorphic function of coordinates. On a Riemann
surface this means that the wave function is holomorphic  in complex (or isothermal) coordinates where the metric is \(ds^2=\sqrt g dzd\bar z\).  Therefore  holomorphic  collective  modes { suffice for a complete field theory of the QHE}. On genus-0
surfaces
they are  power sums  \begin{align}a_{-k}=\sum_{i=1}^N z_i^{k},\quad k\geq1,\quad D\varphi=\prod_{k > 0}
da_{-k} d\bar a_{-k},\nn\end{align} 
The  { sum is taken in the $N\rightarrow \infty$ limit} and the measure of integration \(D\varphi \) represents a functional integration over the real   {\it collective field } $\varphi(\xi)\), where we denote
\(\xi=(z,\bar z)\). { For further discussion of the measure, see Sec.(6). } The field is defined such that its current, the holomorphic derivative $\partial _z\varphi$,  is a generating function of the modes $a_{-k}$ 
    \begin{align}\la{5}
\ii\p_z\varphi \equiv
 -\ii\sum_{k\geq 1}a_{-k} z^{-k-1}.
\end{align} 
In this definition we assume that the field  has no zero modes \(\int
\varphi\, dV=0\) and is therefore globally defined on the Riemann surface. Expectation values are    obtained by a functional integral over the field with the appropriate action\begin{align}
\<\mathcal{O}\>=\frac{\int\mathcal{O}[\varphi] e^{-\Gamma[\varphi]} D\varphi}{\int e^{-\Gamma[\varphi]}
D\varphi}\la{O}
\end{align}  
as opposed to  the multiple integral in \eq{2}.  
 The collective field $\varphi$ defined by its   expansion at infinity \eq{5} can be extended to the finite part of the plane excluding the positions 
 of particles where the current has  poles  \(\p\varphi|_{z\to z_i}\sim-1/(z-z_i)^{}
\).  This field is defined  as \begin{align}\la{14}
\varphi(\xi)  = 4\pi \sum_iG(\xi,\xi_i),  
 \end{align} 
where $G$ is the Green function of the Laplace-Beltrami operator $\Delta$ {with the zero mode removed}, and which satisfies
\begin{align}\nn
-\Delta G(\xi,\xi')=  \delta^{(2)}(\xi-\!\xi')-\frac 1V.\end{align}
 By definition, the collective field is a solution of the Poisson equation \begin{align}\la{15}-\Delta\varphi = 4 \pi (\rho-\frac NV),\end{align}  
  where $\rho(\xi)$ is the particle density. 
 \smallskip

% If the density of particles $\rho$ is a smooth function, we may replace Eq.  \eq{14}  by the integral \(\varphi(\xi)=4\pi \nu^{-1} \int G(\xi,\xi')\rho(\xi') dV'\).  
 
 We now specialize our discussion to the Laughlin state on genus-0 surfaces,
but the final results  hold for any genus. 
% First, we recall some basic aspects of FQH states on a compact surface
%with  a metric \(g\). For more details, see \cite{CLWBig}.  For simplicity
%we consider surfaces of  genus zero. 
The Laughlin wave function  reads
 \begin{align}\label{wfu}
 \Psi & = \frac{1}{\sqrt{\mathcal{Z}}}\prod_{i<j} (z_i-z_j)^{m}  e^{
\frac 12 
\sum_{i} Q(\xi_i)}, \\&\hbar \Delta Q=-2 e B,\label{wfu1}
\end{align}

where $m = 1/\nu$ is an integer, $\nu$ is the filling
fraction,   and \(Q\) is the `magnetic' potential of a slow varying magnetic field $B$.  Below we set \(e=\hbar=1\). 

The normalization $\mathcal{Z}$, known as the generating functional, was
studied  in \cite{clw,CLWBig}. The generating functional is  independent
of the choice of coordinates and depends only on the geometry  of the surface  through functionals of the metric.
 %$B\sqrt g=2\ii(\bar \p A-\p \bar
%A)$.
% The normalization $\mathcal Z[g]$ encodes the structure of the underlying
%geometry. We call it the {\it generating functional}\begin{align}\la{genf}
% \mathcal Z [g]= \int \prod_{i<j} |z_i-z_j|^{2\beta}\prod_{i=1}^N e^{- N_\Phi
%K (\xi_i) }dV_i,  \end{align}
%where $dV_i=\sqrt{g(\xi_i)\,}d^2\xi_i$. 

%The integral converges if the number of particles $N \leq \nu N_\Phi + 
%1$  (more generally,  $N \leq \nu N_\Phi +  \frac\chi 2$, where $\chi$ is
%the Euler characteristic of the manifold {\blue [For higher genus surfaces,
%integral is always convergent - fundamental domain is finite]}). In the %genus zero case, this condition ensures that the wave function is    regular
%at infinity.

 At a given magnetic field the state is normalizable if the  maximal number { of particles is} \begin{align}\la{91}N = \nu N_\phi +\frac 12  \chi,  \end{align}
  where \(\chi\) is the Euler characteristic of the surface (\(\chi=2\)
  for a sphere) and  \(N_\phi=\frac {1}{2 \pi} \int B\,dV  \) is the total number of
  magnetic flux quanta. We assume that the state contains a maximal number of particles so the surface is completely filled and  the particle density
has no boundary.

Our goal is to represent the probability density  $dP = | \Psi|^{2} \prod_{i}
dV_{i} $ as a functional integral over the collective field Eq.\eq{14} such
that $dP \to e^{ - \Gamma[\varphi]} D \varphi$.

\smallskip

\paragraph{Main Results}
 Now we can formulate some    results for the Laughlin  state. We compute the action  $\Gamma[\varphi]$ in \eq{O} in the leading $1/N$ approximation. The  action consists of  three  parts  \be\Gamma[\varphi] =  \Gamma_{G}[\varphi] + \Gamma_{ B}[\varphi] +\Gamma_{L}[\varphi]\la{2711}\ee which are  conveniently written
in terms of the field \(\varphi\) and related field   $\sigma=\log\sqrt {\rho/(N/V) } $ \begin{align}
 \Gamma_{G}[\varphi] & =  \frac{1 }{8\pi\nu} \int  \left[  (\nabla \varphi)^{2} -  R \varphi -4 \nu B\varphi \right]dV,\la{gammaG}\\
\Gamma_{ B}[\varphi]& =\frac 2{\nu} \left( \nu- \frac 12\right) \frac NV \int e^{2\sigma} \sigma\, dV,\la{gammaS}\\
\Gamma_{L}[\varphi] &=\frac{1}{24\pi} \int \left[(\nabla \sigma)^{2} + R \sigma \right]dV.\la{gammaL}
\end{align}
%\begin{align}
%\!\!\!\Gamma[\varphi]\!\!&=\!\frac 1\pi\int\!\Big( \frac 1{8}\left[\nu(\nabla\varphi)^2\!- %\varphi R \right]\,\!\!\!-\!2\pi (1-\frac{1}{2\nu})\bar\rho\,\,e^{2\sigma}\sigma %\, \!+\!   \frac{1}{24} \left[(\nabla \sigma)^2 \!+\!R \sigma \,\right]\Big)\!dV.\la{2710}
%\end{align} 
  where \(R\) is a scalar curvature of the surface.  The actions (\ref{gammaG}-\ref{gammaL}) are derived in sections 4-7. We remind that the field
\(\varphi\) is defined such that \(\int \varphi\,dV=0\), so the coupling
with the curvature \(R\) { and magnetic field $B$} in \eq{gammaG} occurs only if the curvature { and magnetic field}
are not uniform.   If they are uniform, the magnetic field enters only through
relation \eq{91}.

% Another suggestive form of the action is written in terms of the field \(\phi=\varphi-\sigma\) \begin{align}
% \Gamma&=  \frac{1 }{8\pi\nu} \int  \left[  (\nabla \phi)^{2} -  
% R \phi \right]dV+2\bar{\rho}\int e^{2\sigma}
% \sigma\, dV+\\&+\frac{c_H}{24\pi} \int \left[(\nabla \sigma)^{2} + R
% \sigma \right]dV,
% \end{align}where \begin{align}
% c_H=1-3\nu^{-1}.
% \end{align}
The action is non-linear since $\sigma$ and $\varphi$ are connected by the Eq. \eq{15}. It consists  of three distinct terms  at different orders in $1/N$, in descending order. This can be seen by noticing that $\varphi$  defined by \eq{14} is of the order $N$, while $\sigma$ is of the order 1.

The leading term  \eq{gammaG}  of the action is the Gaussian free field with a {\it background charge} which describes the   coupling to curvature, cf.  \cite{Forrester1999,Makarov2011,Zabrodin2006} The
background charge  is directly related to the shift
\(\chi/2\) in \eq{91}.
 Perturbatively,  the action  \eq{gammaG} is equivalent to  the Liouville theory of gravity (see e.g., \cite{FZZ}) in the sense  that  the background charge  increases  the central charge of the Gaussian field  from  \(1\) to $1+3\nu^{-1}.$ As a consequence   the conformal  dimension  of the vertex operator $e^{-a\varphi}$ is 
\begin{align}\la{c}
h_a=\frac 12 a(1-a\nu).
\end{align} 
The conformal dimension is equal to the spin of the quasi-hole. This result  refines the erroneous   notion that the spin of a quasi-hole  matches its mutual statistics and the charge deficit,  both equal
 the filling fraction $\nu$ at $a=1$.\footnote{ To the best of our knowledge the spin of the quasi-hole was correctly computed in \cite{Li92}, see also \cite{Kvorning} and \cite{WZerr}.}

Formally the action  \eq{gammaG} is that of a Gaussian free field and possesses conformal invariance. This invariance  breaks at the next order of the action \eq{gammaS}, except  in the case of the Bosonic Laughlin state \(\nu=1/2\) at which \eq{gammaS} vanishes.

Finally, the Polyakov-Liouville action \eq{gammaL}  manifests  the gravitational anomaly. This part of the action alone  is identical to the action of the  Liouville
  theory of
gravity if the density   $\rho=(N/V) e^{2\sigma}$ is identified as a random
metric (from this point of view, the field $\varphi$ plays the role of
a random K\"ahler potential (cf.\cite{Klevtsov-Mabuchi})).
 The
action does not posses the cosmological term 
since the number of particles is fixed and $\int e^{2\sigma} dV=V$.

%Higher order terms of the action \eq{gammaS} and \eq{gammaL} are perturbations of the conformal part of the action. They are responsible for more subtle effects. 

We can check the consistency  of the action against some known results. 

Minimizing the action we find  the  first three  leading  terms of  the $1/N$ expansion of  the ground state value of the particle density previously obtained in \cite{clw}. If  the magnetic field is uniform
it is also a gradient expansion in curvature  \begin{align}\la{dc}
\!\<\rho\>\!=\!\bar \rho\!+\!\!\left[\!\frac{1}{2\nu}\!\left(\!\nu\!-{\frac
12}\!\right)\! \!+\!\frac{1}{12}\!\right]\!(l^2\!\Delta\!)\frac{R}{8\pi}, \quad \bar\rho\!=\!\frac {\nu B}{2\pi}\!+\! \frac  { R}{8\pi},\!\end{align} 
 where    \(l=\sqrt{\hbar/eB}\)  is the magnetic length.  
 
   The $ \bar \rho$ term in \eq{dc} comes from  \eq{gammaG}.  Integrating over the density yields the particle number \eq{91}, where the $R/(8\pi)$ term yields the background charge of $\chi/2$ due to the Gauss-Bonnet theorem $\int RdV = 4 \pi \chi$.  The order $l^2$ term in \eq{dc}, { which receives contributions from both}
 \eq{gammaS} and \eq{gammaL}
 does not contribute { to the particle number}. 
 
 Linearizing the action on a flat space { yields} the propagator of density  modes  \begin{align}\la{S}\Gamma[\varphi]\approx \frac V{2N}\sum_k  S^{-1}(k)|\rho_{k}|^2,\end{align} where
\(S(k)\) is the   static structure factor  { expanded} to order $k^6$, first computed
in \cite{Kalinay2000}  (see also \cite{clw})    
\begin{align}
S^{-1}(k)= \frac{2}{\nu(kl)^2 }\left(\nu\!+\! (\nu \!-\!\frac 12) (kl)^2 \!+\! \frac{1}{48}(kl)^4 \dots\right) \la{1600}
 \end{align}
 
Other results are described below.
\smallskip

\paragraph{Boltzmann weight }\label{sec_boltzman}
{ The first step in constructing the collective field theory is expressing} the wave function \eq{wfu}  as a functional of the collective field. The amplitude of \eq{wfu}
 is interpreted as the Boltzmann weight of the neutralized Coulomb plasma \(|\Psi|^2\sim e^{-E}\),  with temperature set to unity. We express the energy in terms of the Green function and the K\"ahler potential $K$ defined by the conditions $\p_z\p_{\bar z} K = (\pi/V)\sqrt g$ and $K\sim \log |z|^{2}+\mathcal{O}(1/|z|)$ at infinity. Note that for constant $B$, the potential becomes  $Q= - N_{\phi} K$.  The energy reads
 \begin{align}\nn E = &-2 \int\!\!\int  \rho(\xi) G(\xi, \xi') B(\xi') dV_{\xi} dV_{\xi'}-   N\int Q \frac{dV}V \\&- \frac{1}{2}N N_{\phi}\int K \frac {dV}{V} + \frac{2\pi}\nu \sum_{i\neq j}G(\xi_i,\xi_j).\la{E} \end{align} 
%The amplitude in \eq{13} can be written 
%\begin{align}\la{wf_fields}
%|\Psi|^2=C[g] e^{- E},
%\end{align}
%where $e^{-  E}$ can be interpretted as the Boltzmann weight of a plasma with energy  \( E=(2\pi/\nu) \sum_{i\neq j}G(\xi_i,\xi_j) \). In this interpretation, the temperature is set to one.
The last term in \eq{E}  takes into account the discreteness of particles.  

In the continuum limit, we have to replace the sums over { particle positions}  \(  \sum_{i\neq
j}G(\xi_i,\xi_j)\) by integrals
 over the density taking into account the excluded  self-interaction  at  \(i=j\).    { We must therefore regularize} Green function $G(\xi_i,\xi_j) $ at { coinciding}
points. { The regularized Green function} is  defined by subtracting
the logarithm of  the geodesic distance \( |\xi - \xi'|  g^{1/4}\) between the points in units of the
typical separation between particles, which is of the order of     \( \rho^{-1/2}\)
   \begin{align}\la{19}
G_{R}(\xi) = \lim_{\xi \to \xi'} \left( G(\xi, \xi') + \frac{1}{4\pi} \log[|\xi - \xi'|^2\rho \sqrt g].\right)
\end{align}
 %The latter, however, diverges as a logarithm of the distance
Thus
\(\sum_{i\neq
j}G(\xi_i,\xi_j)\) must be replaced by \begin{align}  
    \int\left[\int G(\xi,\xi')\rho(\xi')dV_{\xi'}
-G_R(\xi)\right] \rho(\xi)dV_\xi.\nn
 \end{align} 
Bringing all pieces  together and integrating by parts 
\begin{align}\la{energy}
  E=  { E_{0}}+\Gamma_G[\varphi] -\frac{1}{ 2\nu} \int\rho\log\rho\,
dV,
\end{align}
where $\Gamma_G[\varphi]$ is given by \eq{gammaG},   and 
 \begin{align}
 E_0
 %&= -\bar\rho\int\left( Q + \frac{N}{2\nu } K+\frac{1}{2\nu} \log\sqrt {g}\right) dV\\
  =\frac{N}{ \nu V} \int \!\!\int \log |\xi - \xi'|^{2} \left(\bar \rho(\xi')-\frac 12\frac N{ V}\right)dV_\xi dV_{\xi'}\nn
%&- \frac{1}{2\nu} \int \int \bar{\rho}^{2} \log |z - z'|^{2} dV dV'\la{E0}
\end{align}
where   \(\bar\rho\) is defined in \eq{dc}.
%\begin{align}
%\varphi_{0} &= Q + \frac{1}{2\nu} \log \sqrt{g} + \frac{N}{2\nu} K\\
%- \Delta_{g} \varphi_{0} & = 4 \pi \nu^{-1} (\frac{\nu}{2\pi} B + \frac{1}{8\pi } R -  \frac{1}{2}\bar{\rho} )\\
%\varphi_{0} &= -\frac{1}{4\pi} \int \log |z - z'|^{2} \left( -\Delta \varphi_{0} \right) dV\\
%& =  - \int \log |z - z'|^{2} \left(\frac{1}{2\pi} B + \frac{\nu^{-1}}{8\pi} R - \frac{\nu^{-1}}{2} \bar{\rho}\right)\\
%E_{0} &= - \int \bar{\rho} \varphi_{0} dV\\
%& =\frac{\bar{\rho}}{2\pi \nu} \int \int \log |z - z'|^{2}  \left( B + \frac{1}{4} R\right) dV dV'\\
%& - \frac{1}{2\nu} \int \int \bar{\rho}^{2} \log |z - z'|^{2} dV dV'
%\end{align}
This gives the field theoretical representation of the wave function.   We comment that the short  distance regularization is determined by the density $\rho$ and for that reason depends on the state of the plasma. A similar regularization scheme was employed for a 1D plasma in Ref.\cite{Dyson}.
 \smallskip

\paragraph{Entropy}\label{sec_entropy}The next step is to pass from integration
over coordinates of individual particles to integration over the macroscopic
density. This is a standard { method}    in statistical mechanics (used in a
setting  similar  to ours in \cite{Dyson}).
% It is also known in the mathematical literature as Sanov's theorem (see %e.g. \cite{berman2011kahler}). 
The transformation defines the Boltzmann entropy 
\(S_B[\rho]=-\int
\rho\log(\rho/\bar{\rho})\, dV\)  
    \begin{align}\prod_i  \sqrt{g(\xi_i)}d^2\xi_i\to
e^{S_B}D\rho.\nn \end{align}
Combining the Boltzmann weight and the entropy together we obtain the probability density \begin{align}\nn
dP\to  e^{- E[\rho]+S_{B}[\rho]}D\rho.
\end{align}
Here, the free energy of local equilibrium is\begin{align}
 {E-S_B= E_{0} + \Gamma_G +\Gamma_{ B}. }\nn
\end{align}
We observe that the Boltzmann entropy and the  short distance  regularization of the Coulomb energy \eq{energy} combine to form $\Gamma_{ B}$.  % 
% \footnote{ {\blue Another symmetry of the action. It can be written almost entirely in terms of the Mabuchi functional. Let us define $\beta E - S = \Gamma'[ \rho, g]$ as a functional of the density $\rho$ and the metric $g$, then
% \begin{align*}
%  \Gamma'[\rho, g]  & = - \frac{\beta N}{4} A_{M}\left[ (V/N) \rho \sqrt{g}, \sqrt{g_0}\right] + \int \rho \log (V\rho/N) dV - \left( \frac{\beta}{2} - 1\right) N \log (N/V)
% \end{align*}
% where the mabuchi functional $A_{M}[g_{1}, g_{2}]$ is a functional of two metrics $g_{1} = e^{2\sigma} g_2$. 
% The second term (Boltzmann entropy) is responsible for breaking conformal symmetry. Specifically, 
% \begin{align}
% \Gamma[ \rho e^{- 2\sigma} , e^{2\sigma} g ] = \Gamma[\rho, g]  - 2  \int \sigma \rho dV
% \end{align}
% 
% It seems that we have two options - 1) expand around classical solution and determine action, or 2) add corrections to action which leave the action invariant under some transformation of $\rho$ (or $\varphi$) and $g$. What is the correct way to proceed with 2)?
% 
% }}
\smallskip

\paragraph{  Ghosts}
The next step is to determine the measure \(D\rho\). 
  Passing from $\rho \to  \varphi$ comes at the price of a Jacobian, which is given by the spectral determinant of the Laplace-Beltrami operator 
 \begin{align}\la{23}D\rho\sim  {\rm Det} ( - \Delta)D\varphi.\end{align}The determinant can be represented by \((1,0)\) Faddeev-Popov ghosts as     \( {\rm Det} ( - \Delta)=\int e^{-\int\bar\eta
(-\Delta)\eta\, dV} D\eta
D\bar\eta \), where     \(\eta\) are complex fermionic modes. 
\smallskip

%The full action then consists of the Coulomb action \eq{21} plus a ghost contribution \begin{align} \Gamma_{C}[\varphi]+\log{\rm Det} ( - \Delta_{g'}).\la{coulomb_ghost}
%\end{align} 
 %Thus the the probability density becomes\begin{align}\la{P1}
%dP=\mathcal{Z}[g]^{-1}e^{ -2\pi N N_{\sigma} A^{(2)}[g]-\frac{1}{4}\beta NA_M^{}[g]
%}  e^{-\Gamma_C[\varphi]-\Gamma_{\rm
%ghosts}[\eta]}D\varphi  D\eta
%D\bar\eta. 
%\end{align}
\paragraph{Gravitational anomaly}\label{sec_grav} The last  step involves the functional measure in \eq{23}. The procedure we outline below is commonly used in the theory
of quantum gravity. Let us denote by \(X\) a field  \(\varphi\) or
  ghosts \(\eta,\bar\eta\) and consider the deviation  \(\delta X\) from a given value
of the field, say its mean. We define the norm of  the deviation
 as   \begin{align}
|| \delta X ||^{2} =  \sum_{i = 1}^{N} (\delta X(\xi_{i}) )^{2} = \int (\delta
X)^{2} \rho dV\la{24}
\end{align} and assume that the measure is  normalized  as   $\int D X \exp [- || \delta X||^{2} ]= 1$. Such  normalization  is supported
by calculations based on the Ward identity { for Laughin states} \cite{CLWBig}. \ Thus the measure for both $\varphi$ and the ghost fields  depends in a nontrivial fashion on the density, and thus on $\varphi$ itself. So although the ghosts appear decoupled from the rest of the action, in fact they are not. 

The density $\rho$  appearing in \eq{24} can be treated as a conformal factor  of the metric and  thus  removed  from the measure by a conformal transformation of  coordinates \(dV\to\rho^{-1}
dV\,\). It is known, however,  that under
conformal transformation the measure transforms anomalously  as \begin{align}
D_{} X \to e^{  c_{X}\, \Gamma_{L}[\sigma]} DX, \nn
\end{align}
where $c_{X}$ is the central charge of the field $X$, where \(\Gamma_L[\rho]\)
is the Polyakov-Liouville action \eq{15} \cite{Polyakov1987}, see also \cite{DistlerKawai}. This is the {\it Weyl} or {\it gravitational anomaly} { which appears here} in a similar fashion as in the quantum
theory of  gravity. Applying this to the collective field \(\varphi\) with the central charge
\(+1\) and ghost with the central charge \(-2\)  we obtain the measure %{\blue [footnote]We adopt  the  measure of  integration  $D\varphi$ rather than $D\log\rho$ as commonly used in the theory of quantum gravity.} 
 \begin{align}\nn
 e^{ - \Gamma_{L}[\rho]} D \varphi  D \eta D \bar{\eta.}
\end{align}
   After the Polyakov-Liouville action is taken into account the short distance
regularization of the field \(\varphi\) and ghosts does not depend on
density. Since the ghosts are decoupled  their contribution is
the spectral determinant
of the Laplace operator. Summing up, the probability
distribution is  \begin{align}
 dP=\mathcal{Z}^{-1}{\rm\ Det}(-\Delta)e^{- E_{0}-\Gamma[\varphi]}D\varphi.\la{25}
\end{align} The ghosts determinant  contributes to 
the finite size correction to the { free energy of the Coulomb plasma \cite{Jancovici1, CLWBig}}. 

\smallskip

 Now we turn to some applications. 

\smallskip

\begin{comment}
Summing all of the pieces together, we obtain   the generating
functional as the functional integral\begin{align}
\mathcal{Z}[g]= e^{-2\pi N N_{\sigma} A^{(2)}[g]+\frac 1{4\nu} NA_M[g]}{\rm Det}
\left(-\Delta_{g}\right)
\int  e^{-\Gamma[\varphi]}D\varphi,
\la{311}\end{align}
with the action  given by \eq{2710}.
\end{comment}
%\begin{align}
%\!\!\!\Gamma\!\!=\!\int\!\left(
%\frac 1{8\pi\beta}\left[(\nabla\varphi)^2\!-\!\beta  \varphi R \right]\,\!\!\!-\frac
%12\!\left(\beta
%-\!2\right)\rho\log\rho\,
%\!+\!   \frac{1}{96\pi} \left[(\nabla \log \rho)^2 \!+\!2R\log \rho \,\right]\right)\!dV\la{271}
%\end{align} 

\paragraph{Density and generating functional}

We start
from computing the generating functional - { the normalization factor of the Laughlin wave function
or \eq{25}}.  

The integral of the lhs of \eq{25} is 1. The relevant contribution to the
 integral of the  rhs of \eq{25} comes from  the Gaussian approximation.
It consists of the on-shell action \(\Gamma[\varphi_{c}]\) computed on the ``classical" solution $ \varphi_{c} $, { which minimizes the action}. Computing
 Gaussian fluctuations it is sufficient to take into account  only the leading part of  the
action \eq{gammaG} \be\nn\int e^{-\Gamma[\varphi]}D\varphi=[{\rm\ Det}(-\Delta)]^{-\frac 12}e^{-\Gamma[\varphi_c]}.\ee  Thus integrating \eq{25} gives
\begin{align}
\mathcal{Z}=[{\rm\ Det}(-\Delta)]^{\frac
12}e^{- \Gamma_{0}},\quad \Gamma_0=E_0+\Gamma[\varphi_c].  \la{261}
\end{align} 
%where we denote \(\Gamma_0=E_0+\Gamma[\varphi_c]\). 

In the three first  leading orders in \(1/N\)
solution of    \(\delta\Gamma[\varphi]/{\delta\varphi}=0\) is the ground state value of the field \(\varphi_c=\<\varphi\>,\) which, through \eq{15}
determines  the
ground state value of   the density.  Solving
in the leading order in \(1/N\) we obtain Eq.\eq{dc}.

Inserting \eq{dc} back into \eq{2711} we find   \be\Gamma[\varphi_c]=-\frac
{2\pi}{\nu }\int \int \bar\rho(\xi') G(\xi, \xi') \bar\rho(\xi') dV_{\xi} dV_{\xi'}.\nn\ee
% where the source   $J = 2 \nu B + R/2$. 
% to be equal   \be \frac{1}{8\pi \nu }\int
%[ \frac{2 e}{\hbar}\nu B(\xi) + \frac{1}{2} R(\xi)] G(\xi,\xi') [\frac{2e}{\hbar} \nu B(\xi') + \frac 12R(\xi')] dV_\xi dV_{\xi'}.\nn\ee 
%Combining this result  with \(E_0\) in \eq{119} we obtain  \(\Gamma_0\) in \eq{261}. 
The final result for the functional { \(\Gamma_{0}\)}  in \eq{261} is best expressed in terms of    the gauge potential
 and spin connection. Their complex components are  defined by \be 2\ii(\p_{\bar z}A_z-\p_z A_{\bar z})= { B\sqrt g},\quad 2\ii(\p_{\bar
z}\omega_z-\p_z \omega_{\bar z})=\frac1 2 R\sqrt g.\nn\ee In the  transverse
gauge \(\p_{\bar z}A_z=-\p_z A_{\bar z},\;\p_{\bar z}\omega_z=-\p_z
 \omega_{\bar z}\)  the functional $\Gamma_0$ has a compact form
   \begin{align}
 \Gamma_{0}=-\frac{2}{\pi\nu}\int\left|\left (\nu A_{ z}+ \frac 12\omega_{ z}\right)\right|^2dzd\bar z.\nn
\end{align} 

It  remains  to recall the value of the spectral determinant  of the Laplace operator in \eq{261}. Up to a metric independent terms it is  given by the Polyakov formula \cite{Polyakov1981}
$$\log {\rm Det}(-\Delta)=-\frac 1{3\pi}\int |\omega_z|^2 dzd\bar z.$$ 
As a result (cf.,\cite{CLWBig}) \begin{align}
\log\mathcal{Z}\!=\!\!\int\left[\frac{2}{\pi\nu}\left|\left (\nu A_{ z}+ \frac 12\omega_{
z}\right)\right|^2\!\! { -\frac
1{6\pi}}|\omega_z|^2 \right]dzd\bar z.\la{28}
\end{align} 
In the form \eq{28} it is valid on a surface with any genus.  
 
The authors of Ref.\cite{KW2015} argued that the elements of the Hessian matrix of the  generating functional 
 \be\sigma_H=\frac\pi 2\frac{\delta^2\log \mathcal{Z}}{\delta A_z\delta A_{\bar z}}, \; 2\varsigma_H=\frac\pi
2\frac{\delta^2\log \mathcal{Z}}{\delta \omega_z\delta A_{\bar z}},\;-\frac{c_H}{12}=\frac
\pi 2 \frac{\delta^2\log \mathcal{Z}}{\delta \omega_z\delta \omega_{\bar z}}\nn\ee are  universal transport coefficients precisely quantized on QH-plateaus.  Here $\sigma_H$ is the Hall conductance,  $\varsigma_H$ determines the current caused by changing of the metric and the third coefficient, $c_H$,   describes forces exerted on the fluid as a result of a changing  the metric. We refer to \cite{KW2015} for further details. For Laughlin states these coefficients are encoded in \eq{28}
\be\sigma_H=\nu,\quad \varsigma_H=1/4,\quad
c_H=1-3/\nu \label{TK}
\ee

\smallskip

\paragraph{Quasi-holes - gauge anomaly.  } Introduced by Laughlin \cite{Laughlin1983}, a quasi-hole state with charge $a$ on a compact surface reads
 \begin{align}\la{psi}
\Psi_a  \!=\!\frac {e^{ \frac{1}{2}\nu a [ Q(w) - a K(w)]  }}{\sqrt{\mathcal{Z}_a[w,\bar w]}}\left[\prod_{i = 1}^{N} (z_{i} \!-\! w)^{a}e^{ \!- \!\frac{a}{2}  K(z_i,\bar z_i) } \right]\Psi,
\end{align} 
where \(w\) is a  holomorphic coordinate of the quasi-hole, $\Psi$ is the ground state  \eq{wfu}  with $N$ particles subject to the condition  \eq{91}, $a$ is a positive integer less than $m=1/\nu$, and $K$ is defined above \eq{E}.  The factor of $\exp\left(- \frac{a}{2} K(\xi_i)\right)  $ neutralizes the insertion
of the quasi-hole. This state covers the entire surface. { The exponential factor of \( \frac{a \nu }{2}  [Q - a K  ]\) } in \eq{psi} is added for a convenience.

A quasi-hole is represented by the vertex operator \(V_a(w,\bar w)=e^{-a\varphi(w,\bar
w)}\).
In particular
the normalization factor \(\mathcal Z_a\), the generating
functional for a quasi-hole state, reads up to constants \begin{align}
\mathcal{Z}_a[w,\bar w]\sim \Big\langle
V_a(w,\bar w)\Big\rangle,\nn\end{align} where the average  is taken over the
ground  state \eq{wfu} without the quasi-hole.  As such  the quasi-hole may be seen
as  a source
for the action  \eq{gammaG} {
 \(\Gamma\to \Gamma + a\varphi(w)\). }   However, there is a caveat. The quasi-hole   disturbs the electronic density around itself  in a vicinity of the size
of magnetic length.  At the limit of a vanishing magnetic  length the density
becomes singular. At the same time the derivation of the    action  was based on    the    assumption
that the density is smooth. Therefore the derivation must be reexamined   to take into account the feedback of the singularity. 
    
The leading \(1/N\) value of \eq{28} is given  by the Gaussian
part of the action \eq{gammaG}   
 \be\mathcal{Z}_a\approx  \exp\left(-\!a\<\varphi \>\!+\!\frac {a^2} 2\<\varphi^2\>_c\right).\la{251}\ee
The mean of the field \(\varphi\) determined by \eq{gammaG} is $$\<\varphi(\xi)\>\approx 4\pi\int G(\xi,\xi') \bar\rho(\xi')
dV_{\xi'}=\nu Q + \frac 12\log\sqrt {g(\xi)},$$
the variance is 
 \( \<\varphi^2\>_c\equiv\<\varphi^2\>-\<\varphi\>^2=4\pi \nu G_R\), where the regularized Green function
 is given by \eq{19}. But the \(G_R\) depends on the  density
itself,  and in the leading approximation  one replaces the density by its { mean such that} \(\<\varphi^2\>_c=\nu\log\left(\<\rho\>\sqrt g\right)\). Putting this together we obtain  \begin{align}\la{29}
\mathcal{Z}_a \approx\left(\sqrt{\<\rho\>}\right)^{\nu a^2}  \left(\sqrt{g}\right)^{- h_{a}},
\end{align}
where $h_{a} =\frac{a  }{2}(1 - \nu a )$ { is the conformal  dimension} as in \eq{c}.

 In the leading approximation the factor \(\<\rho\>\) in \eq{29} can be treated
as a constant. 
Then \eq{25}  suggests that \(h_a\) is the conformal dimension of the quasi-hole
state: 
the quasi-hole state transforms  as a primary field under a holomorphic  transformation.  Symbolically  
$$w\to f(w),\quad V_{a}\to  (f'(w))^{h_{a}} V_{a}$$

Because the state is holomorphic {(up to the normalization factors in \eq{psi})}the holomorphic dimension \(h_a\) is also
the spin of the state. Later we show this in a more direct manner.

In the next to the leading approximation we cannot { assume}  the density is \eq{29} to be a  { constant}. { As with the} gravitational anomaly
above, { the field transforms as} \(\varphi\to \varphi-a\nu\log\sqrt\rho\), { which modifies}   the vertex operator  \begin{align}
V_a=(\sqrt\rho)^{\nu a^2}
e^{-a\varphi},\nn
\end{align}  such that the regularization of the {two-point correlation function} at { coincident}
points is independ on the state density.  Alternatively, we may say that the quasi-hole contributes to the action as  a source { \(\Gamma\to \Gamma+ a\varphi-a^2\nu\log\sqrt\rho\).}
Thus the stationary point
of the action reads\begin{align}
\frac{\delta \Gamma}{\delta\varphi(\xi)}=-a\left(1+\frac {\nu a}{8 \pi\rho}\Delta\right)\delta(w-\xi).\la{32}
\end{align} 
In the linear approximation we treat \(\rho\) in \eq{32}
as a constant \(\approx\nu/(2\pi l^2)\) and use \eq{S}. As a result we obtain the first two terms of the expansion in \((kl)^2\) \begin{align}
\rho_k\!\approx \!  \frac{2  \nu a}{(kl)^{2}}  \left(\!-1\!+\frac {a}{4}(kl)^2\right) \!S(k)\!\approx\!\!-\nu a\!+\!\frac
{(kl)^2}2(a\nu\!-\! h_a)\nn.
\end{align}
Equivalently the first two moments of the density  $\delta \rho = \langle 
\rho \rangle - \frac NV$ are

 \begin{align}
&m_0=\int   \delta \rho\,  d{V} = - \nu a,\label{C}\\&
m_2=\frac 1{2l^2}\int \,  r^2\delta \rho\, d{V}  =-\nu a+ \frac{1  }{2}a(1 - \nu a ).\label{D}
\end{align}
The first  formula describes the    fractional
  charge deficit $-\nu a$. This  result goes back to \cite{Laughlin1983}.
The  second moment  is more involved  \cite{Samaj2007,Jancovici2008,CLWBig}. Curiously, the second moment vanishes
at $\nu  =\frac  1 3$ and $a = 1$.

 { Having determined the generating functional, we compute the adiabatic phase \(\gamma_\mathcal{C}\)  acquired by the  quasi-holes by transporting one around a closed path \(\mathcal{C}\). }
 
  For simplicity we compute the adiabatic phase when
one hole   with coordinate \(w_1\) moves around a closed path \(\mathcal{C}\)   enclosing  another quasi-hole
with coordinate \(w_2\).
 The extension of  (\ref{251},\ref{29}) to the case of two quasi-holes
is \be\mathcal{Z}_{a_1a_2}(w_1,w_2)=\mathcal{Z}_{a_1}(w_1)\mathcal{Z}_{a_2}(w_2)  e^{4\pi  \nu   a_2a_1G(w_1,w_2)} ,\la{31}\ee
where we used  \(\langle \varphi(w_{2})\varphi(w_{2}) \rangle_{c} = 4\pi \nu G(w_1,w_2)\)
and \eq{251}. 

The adiabatic phase
reads \begin{align}
\gamma_\mathcal{C}=2 \ii \int\left[ \oint_\mathcal{C}\overline{\Psi}\partial_{w_1}\Psi
dw_1\right]dV_1\dots dV_N&.\nn
\end{align} 
%= 2 \ii \oint_\mathcal{C}\langle
%\{w_k\}|\partial_{w_1}|\{w_k\}\rangle dw_1\).
Since the state is a holomorphic function of  position of the quasi-holes, only normalization factor in \eq{psi} contributes to  the phase
\begin{align}\gamma_\mathcal{C}=-2\pi{a_1}\nu\,\Phi_{\mathcal{C}}+\ii \oint_\mathcal{C}\partial_{w_1}\log\mathcal{Z
}_{a_1a_2}dw_1. \nn
\end{align}
The first term is  the Aharonov-Bohm phase picked up by a particle with charge
$-a_1\nu$  enclosing 
 the magnetic flux  \(\Phi_{\mathcal{C}}=( N_\Phi+a_1+a_2)\text{Area}(\mathcal
 C)/V\) in units of the flux quantum.
 The contribution of the second term follows from \eq{31}    
 \begin{align}\la{Berryphase}
\ii \oint_\mathcal{C}\partial_{w_1}\log\mathcal{Z
}_{a_1a_2}dw_1= -
%+\frac{\mathcal{S}+4s}{8\pi}(R-R_0)\right) dV
h_{a_1}\Omega_\mathcal{C}+2\pi \nu a_1 a_2.\end{align}
It contains  the solid angle \(\Omega_\mathcal{C}=\ii\oint d\log\sqrt g =\frac
12\int _\mathcal{C}R dV\). The coefficient in front of it is the   spin  of the quasi-hole, equal to the holomorphic dimension \eq{c}. This formula extends the result of Refs.\cite{Li92}, which was for the adiabatic phase of a single quasi-hole ($a = 1$) on a sphere. 

 The last term in \eq{Berryphase}  \(  4\pi  \ii  \nu a_2
 a_1 \oint dG(w_1,w_2)\), which vanishes if the contour $\mathcal{C}$ does not enclose $w_2$, is commonly referred to as the mutual statistics of the quasi-holes.   When the quasi-holes are identical, it is equal to $\nu a^{2}$, and differs from the spin.

 \paragraph{Effect of spin} Lastly, we comment on the effect of spin of quantum Hall states.  The spin, yet another characterization of the QH state  was introduced in Ref. \cite{CLWBig}. The inclusion of spin comes as a generalization of the lowest Landau level (LLL). We recall that the LLL  are defined as zero modes of the anti-holomorphic component of the kinetic momentum operator  $\bar \pi  = - i\hbar \bar \partial + \hbar s \bar \omega - e \bar A$ where $\bar \omega = - (i /2 )\bar \partial \log \sqrt g$, where parameter  $s$ is the spin. Throughout  the paper we set the spin to zero.  Inclusion of spin effectively shifts the  potential $Q$ in \eqref{wfu1} by $- s \log \sqrt{g}$, such that the modified $Q$ now satisfies the Poisson equation $ \Delta Q=-\frac{2 e}{\hbar} B+s R$. As a result, the   action acquires an additional term  
 %$  s \int \rho \log \sqrt{g} dV =
 $ \frac{s}{4\pi} \int \varphi R dV, $
% -  s \bar{\rho} \int \log \sqrt{g} dV$,
 which shifts the background charge in the Gaussian action 
\begin{align}
 \Gamma_{G}[\varphi] & =  \frac{1 }{8\pi\nu} \int  \left[  (\nabla \varphi)^{2} -  (1-2\nu s)R \varphi -4 \nu B\varphi \right]dV.\nn
% \Gamma_{S}[\varphi] &=\text{Here we write the modified formula}
\end{align} 
%The metric functional $E_{0}$  stays as it is written in \eq{E0}, but with the density appropriately defined as $\bar{\rho} = \frac{\nu B}%{2\pi} + \frac{(1 - 2 \nu s)}{8\pi} R$. 
The Boltzmann entropy \eq{gammaS} and the Polyakov-Liouville action \eqref{gammaL} remain the same. 
Below we list  some effects of spin.
 
Spin does not appear in  local properties evaluated at distances  where change of curvature is negligible, for example in a flat space. In particular the structure factor $S(k)$ \eq{1600}, the charge of the quasi-hole $m_0$  \eqref{C} and its moment  $m_2$ \eqref{D} are independent of spin. 

However geometric characteristics   depend on spin. As such, the relation  \eq{91} between  the total number of particles and magnetic flux   becomes
\begin{align}\la{Nspin}N = \nu  N_\phi +\frac 12(1 - 2 \nu s )   \chi .  \end{align}
The spin  modifies the conformal dimension \eq{c} defined in \eq{29} and appearing in the adiabatic phase \eq{Berryphase} 
\be h_{a} = \frac{1  }{2} a( 1 - 2 \nu s - a\nu).\nn\ee However, the second moment \eq{D}  will  not acquire any spin dependence, and will maintain its relation to the conformal dimension $m_{2}=(1-s)m_0+ h_a$.

Spin also enters the  generating functional \eqref{28}  
 \begin{align}
&\log\mathcal{Z}\!=\!\!\int\left[\frac{2}{\pi\nu}\left|\left (\nu A_{ z}+ \frac 12(1-2\nu s)\omega_{
z}\right)\right|^2\!\! { -\frac
1{6\pi}}|\omega_z|^2 \right]dzd\bar z\nn
\end{align} 
Consequently, the Hall conductance  does not depend on spin, but the geometric transport coefficients  in \eqref{TK} do
 \begin{align}
& \varsigma_H=\frac 1{4}(1-2{\nu s}),\; c_H=1-3\nu^{-1}\left(1-2\nu s\right)^2\nn
\end{align} 
 For more  details regarding the inclusion of spin into the FQHE on a curved space, see \cite{CLWBig}.

\smallskip

\bigskip

\paragraph{Conclusion}
In summary, we formulated the theory of the Laughlin QH-states as a field
theory of a scalar Bose field. The field theory consists of the   Gaussian action with the background charge and the sub-leading  corrections
representing the gravitational anomaly.  We demonstrated  that this theory   captures conformal properties of quasi-holes,  the  adiabatic transport, and  clarifies the effect of the gravitational  anomaly.  
\bigskip

Finally we comment that the action   similar to \eq{2711}
has been considered in \cite{Klevtsov-Mabuchi} as an admissible  
action for a random metric. The actions become analogous upon identifying  the
fluctuating  density as a random metric and the field \(\varphi\) as a fluctuating
K\"ahler potential. %{\blue The formula \eq{23} also appeared in Ref. \cite{Klevtsov-Mabuchi} as  a relation between  the invariant measures on conformal factors and on K\"ahler potentials.} 
We thank S. Klevtsov for bringing Ref. \cite{Klevtsov-Mabuchi} to our attention. 
\bigskip

\paragraph{Acknowledgments}\noindent We thank
A. Gromov, S. Klevtsov  and S. Zelditch  for comments on the draft of the paper.  The work  was supported by NSF DMR-1206648, DMS-1156656,
NSF DMR-MRSEC -1420709

\bibliography{fqhe_curved_refs.bib}
\end{document}